\documentclass{article}
\usepackage{times}
\usepackage{amsfonts}
\usepackage{graphicx}
\usepackage[pdfmark]{hyperref}
\begin{document}
\noindent
{\Large  STAR PRODUCTS ON GENERALISED COMPLEX\\ MANIFOLDS}
\vskip1cm
\noindent
{\bf Jos\'e M. Isidro}\\
Instituto de F\'{\i}sica Corpuscular (CSIC--UVEG), Apartado de Correos 22085,\\ Valencia 46071, Spain\\
{\tt jmisidro@ific.uv.es}
\vskip1cm
\noindent
{\bf Abstract} We regard classical phase space as a generalised complex manifold and analyse the $B$--transformation properties of the $\star$--product of functions. 
The $C^{\star}$--algebra of smooth functions transforms in the expected way, while the $C^{\star}$--algebra of holomorphic functions (when it exists) transforms nontrivially. The $B$--transformed $\star$--product encodes all the properties of phase--space quantum mechanics in the presence of a background magnetic field.

\vskip1cm

\section{Introduction}\label{einfuehrung}

Ever since its inception by Wigner in 1932 \cite{WIGNER},  quantum mechanics on phase space has remained a source of inspiration for mathematical physicists trying to clarify the deep relation between the classical world and the quantum world.  Some milestones along this route were the pioneering articles of refs.  \cite{GROENEWOLD} and \cite{MOYAL}, which culminated in the formal development of deformation quantisation \cite{DQ}.  A central role in this programme is played by the $\star$--product operation on Poisson manifolds \cite{KONTSEVICH}. Detailed references, as well as an account of the current status of quantum mechanics on phase space from different perspectives, can be found in refs. \cite{MAURICE, GOSSON, GMS, ZACHOS}.

Along an apparently unrelated line, impressive advances in field theory and string theory on noncommutative spaces have been made recently; for a nice review see \cite{SZABO}. Although building on  previous work in noncommutative geometry \cite{CONNES, LANDI}, most of these advances were triggered off by the seminal papers of refs. \cite{CDS, SW}. A key element in these noncommutative theories is the presence of a background magnetic field, also called {\it Neveu--Schwarz field}\/ or $B$--field, whose nonvanishing in the region of spacetime under consideration renders the latter noncommutative.

A natural question to pose is, how does phase--space quantum mechanics change in the presence of a $B$--field? Recent breakthroughs in the theory of complex manifolds and symplectic geometry, that go by the name of {\it generalised complex structures}\/ \cite{HITCHIN, GUALTIERI, LINDSTROM}, provide us with powerful tools to find the answer. Clearly one needs to find out how the $\star$--product, which lies at the heart of quantum mechanics on phase space, gets modified under the presence of a background $B$--field. In other words, one needs to know the transformation properties of the $\star$--product under {\it $B$--transformations}. This is best done by regarding classical phase space, usually considered a Poisson or even a symplectic manifold, as generalised complex and then applying a $B$--transformation. The resulting $\star$--product, that we will denote $\star_B$, will encode all the relevant properties of {\it phase--space quantum mechanics in the presence of a background magnetic field}.

\section{The $\star$--product  without $B$--transformations}\label{stellung}

Consider the linear space $\mathbb{R}^{2n}$ endowed with its usual symplectic form $\omega$. In local coordinates $x^{i}$ we have
\begin{equation}
\omega=\frac{1}{2}\omega_{ij}{\rm d}x^{i}\wedge{\rm d}x^j, \qquad i,j=1,\ldots, 2n,
\label{gestalt}
\end{equation}
the Poisson brackets of any two smooth functions $f,g$ on $\mathbb{R}^{2n}$ being 
\begin{equation}
\{f,g\}=\pi^{ij}\partial_if\partial_jg,  \qquad \omega_{ij}\pi^{jl}=\delta_i^l, \quad i,j,l=1, \ldots, 2n.
\label{klammern}
\end{equation}
Assume that the coordinates $x^{j}$ are Darboux, so we may divide them into $q^j$ and $p_j$,  for $j=1, \ldots, n$. In the ordering  $q^j, p_j$ we have the block matrices
\begin{equation}
\omega=\left(\begin{array}{cc}
0&-{\bf 1}_n\\
{\bf 1}_n&0\end{array}\right), \qquad
\pi=\omega^{-1}=\left(\begin{array}{cc}
0&{\bf 1}_n\\
-{\bf 1}_n&0\end{array}\right).
\label{matzxfer}
\end{equation}
Next regard $\mathbb{R}^{2n}$ as $\mathbb{C}^n$, and define
\begin{equation}
z^j:=\frac{1}{\sqrt{2}}\left(q^j+{\rm i}p_j\right), \qquad j=1, \ldots, n.
\label{zzz}
\end{equation}
In the order $\bar z^j, z^j$, the matrices for $\omega$ and $\pi$ now equal
\begin{equation}
\omega={\rm i}\left(\begin{array}{cc}
0&{\bf 1}_n\\
-{\bf 1}_n&0\end{array}\right), \qquad \pi={\rm i}\left(\begin{array}{cc}
0&{\bf 1}_n\\
-{\bf 1}_n&0\end{array}\right).
\label{lytisch}
\end{equation}

Next recall that noncommutative $\mathbb{R}^{2n}$ is defined by the commutators 
\begin{equation}
[\hat x^{i}, \hat x^j]={\rm i}\theta^{ij}{\bf 1},\quad [\hat x^l, \theta^{ij}{\bf 1}]=0, \quad \theta^{ij}=-\theta^{ji}, \quad i,j,l=1, \ldots, 2n.
\label{komm}
\end{equation}
Above, $\theta^{ij}$ is a constant, antisymmetric real tensor, while ${\bf 1}$ and the $\hat x^l$ are operators on infinite--dimensional Hilbert space (we will henceforth drop the carets on top of the $x^l$). Now the pointwise product of functions $f,g$ on commutative $\mathbb{R}^{2n}$ is replaced, on noncommutative $\mathbb{R}^{2n}$, with the (associative, noncommutative) $\star$--product,
\begin{equation}
(f\star g)(x^l):=f(x^l)\,\exp\left(\frac{{\rm i}}{2}\overleftarrow\partial_i\theta^{ij}\overrightarrow\partial_j\right)\,g(x^l).
\label{stern}
\end{equation}
In the coordinates $\bar z^j, z^j$ of eqn. (\ref{zzz}), noncommutative $\mathbb{R}^{2n}$ of (\ref{komm}) can be rewritten as
\begin{equation}
[\bar z^j, z^l]=-\delta^{jl},\quad [z^j, z^l]=0=[\bar z^j, \bar z^l], \quad j,l=1, \ldots, n,
\label{mmok}
\end{equation}
and the (block) matrix expression for the noncommutativity parameter $\theta^{ij}$, in the order $\bar z^j, z^j$, is
\begin{equation}
\theta=\left(\begin{array}{cc}
0&-{\bf 1}_n\\
{\bf 1}_n&0\end{array}\right).
\label{matz}
\end{equation}
Comparing eqns. (\ref{lytisch}) and (\ref{matz}) we can write a well--known relation between the Poisson tensor $\pi$ and the noncommutativity tensor $\theta$:
\begin{equation}
\pi=-{\rm i}\theta.
\label{homm}
\end{equation}
Next reexpressing $f(q,p)$  as $f(\bar z, z)$ and $g(q,p)$ as $g(\bar z, z)$ we arrive at 
\begin{equation}
(f\star g)(\bar z, z)=f(\bar z, z)\,\exp\left(-\frac{1}{2}\overleftarrow\partial\pi\overrightarrow\partial\right)\,g(\bar z, z), \quad f,g\in C^{\infty}(\mathbb{R}^{2n}).
\label{holstern}
\end{equation}
Above, $C^{\infty}(\mathbb{R}^{2n})$ denotes the algebra of smooth functions on $\mathbb{R}^{2n}$. In particular,  the $\star$--product of any two {\it holomorphic}\/ functions $f,g$ equals their pointwise product:
\begin{equation}
(f\star g)(z)=(f\cdot g)(z)=f(z)\cdot g(z), \qquad f,g\in H(\mathbb{C}^n),
\label{devs}
\end{equation}
where $H(\mathbb{C}^n)$ denotes the algebra of holomorphic functions on $\mathbb{C}^n$.

Let us take stock. We have regarded the symplectic space $(\mathbb{R}^{2n},\omega)$ as the complex space $\mathbb{C}^n$, {\it the latter endowed with a complex structure that is compatible with the symplectic structure $\omega$}.  Commutative $C^{\star}$--algebras of functions to place on $\mathbb{R}^{2n}$ and $\mathbb{C}^n$ are $C^{\infty}(\mathbb{R}^{2n})$ and $H(\mathbb{C}^n)$, respectively. Moreover,  $H(\mathbb{C}^n)$ is a proper subalgebra of $C^{\infty}(\mathbb{R}^{2n})$. However, while the $\star$--product (\ref{stern}) deforms $C^{\infty}(\mathbb{R}^{2n})$ into $C^{\star}(\mathbb{R}^{2n})$ (the noncommutative $C^{\star}$--algebra of functions on noncommutative $\mathbb{R}^{2n}$), the subalgebra $H(\mathbb{C}^n)$ remains undeformed and, in particular, commutative. We may picture this deformation graphically as
\begin{equation}
C^{\infty}(\mathbb{R}^{2n})\stackrel{\star}{\longrightarrow}C^{\star}(\mathbb{R}^{2n}),
\label{ccw}
\end{equation}
while it acts as the identity on $H(\mathbb{C}^n)$:
\begin{equation}
H(\mathbb{C}^{n})\stackrel{\star}{\longrightarrow}H^{\star}(\mathbb{C}^{n}):=H(\mathbb{C}^n).
\label{ccwx}
\end{equation}
Above we have formally introduced the notation $H^{\star}(\mathbb{C}^{n})$ for the $C^{\star}$--algebra of holomorphic functions on $\mathbb{C}^n$ with respect to the $\star$--product (\ref{stern}), even if  $H^{\star}(\mathbb{C}^{n})$ equals the commutative $H(\mathbb{C}^n)$. Loosely speaking we may conclude that there is no deformation quantisation of complex manifolds; only Poisson manifolds can be deformed--quantised.

\section{The $B$--transformed $\star$--product}\label{tionen}

We ask whether the conclusion encapsulated by eqn. (\ref{ccwx}) remains valid in the framework of generalised complex geometry. To answer this question we will first regard $\mathbb{R}^{2n}$ as generalised complex. Next we will apply a $B$--transformation to the Poisson tensor $\pi$ and compute the corresponding $\star$--product, that we will denote by $\star_B$. We need to find out if  holomorphic functions will continue to satisfy the invariance property (\ref{devs}), with $\star_B$ replacing $\star$. 

A comment is in order. By what was said above, regarding $\mathbb{R}^{2n}$ as generalised complex of type $k=0$ is equivalent to regarding $\mathbb{C}^n$ as generalised complex of type $k=n$. This apparent contradiction concerning the type $k$ is due to the fact that the complex structure considered above on $\mathbb{C}^n$ was compatible with the symplectic structure $\omega$ placed previously on $\mathbb{R}^{2n}$. Since we are ultimately interested in doing quantum mechanics, we need to adopt the symplectic viewpoint and regard phase space as generalised complex of type $k=0$.

On a linear space such as $\mathbb{R}^{2n}$, any generalised complex structure of type $k=0$ is the $B$--transform of a symplectic structure \cite{GUALTIERI}. This means that any generalised complex structure of type $k=0$ can be written as
\begin{equation}
\left(\begin{array}{cc}
{\bf 1}_{2n}&0\\
-B&{\bf 1}_{2n}\end{array}\right)
\left(\begin{array}{cc}
0&-\omega^{-1}\\
\omega &0\end{array}\right)
\left(\begin{array}{cc}
{\bf 1}_{2n}&0\\
B&{\bf 1}_{2n}\end{array}\right)
=\left(\begin{array}{cc}
-\omega^{-1}B & -\omega^{-1}\\
\omega + B\omega^{-1}B & B\omega^{-1}\end{array}\right),
\label{ppewez}
\end{equation}
for a certain 2--form $B$, which we will hereafter take to be closed.\footnote{In eqn. (\ref{ppewez}),  the decomposition into block matrices is that corresponding to $\mathbb{R}^{2n}$ as generalised complex, {\it i.e.}, it reflects the direct sum  $T^*\mathbb{R}^{2n}\oplus T\mathbb{R}^{2n}$, hence the blocks are $2n\times 2n$. Eqn. (\ref{ppewez}) is the only occurence of such a block decomposition in this paper, all other block matrices being $n\times n$.} Thus the $B$--transformation law for the symplectic form $\omega$ is \cite{MAGNETIC}
\begin{equation}
\omega\longrightarrow\omega_B=\omega + B\omega^{-1}B.
\label{megatrans}
\end{equation}
It follows by ${\rm d}B=0$ that $\omega_B$ is again a symplectic form \cite{MR}. Correspondingly, the Poisson tensor $\pi=\omega^{-1}$ transforms under $B$ as
\begin{equation}
\pi\longrightarrow\pi_B=\omega_B^{-1}=\left(\omega + B\omega^{-1}B\right)^{-1}.
\label{pitrnsz}
\end{equation}
As in ref. \cite{DIRAC} we will consider the case of weak magnetic fields. This allows us to approximate the right--hand side of (\ref{pitrnsz}) as
\begin{equation}
\pi_B=\pi-\pi B\pi B\pi+O(B^4),
\label{trn}
\end{equation}
because $\pi_B\omega_B=1+O(B^4)$. In what remains we will always work in the weak--field approximation. Let the block matrix decomposition for $B$ in the coordinates $\bar z^j, z^j$ be
\begin{equation}
B=\left(\begin{array}{cc}
B_{\bar z\bar z} & B_{\bar z z}\\
B_{z\bar z}& B_{z z}\end{array}\right), \quad B_{\bar z\bar z}^t=-B_{\bar z\bar z}, \quad B_{z z}^t=-B_{zz}, \quad B_{\bar z z}^t=-B_{z\bar z}.
\label{xyzt}
\end{equation}
Above the superscript ${}^t$ stands for {\it transposition}, while the subindices $\bar z, z$ are true matrix indices only when $n=1$; for $n>1$ they only denote block matrix indices. By eqn. (\ref{trn})
\begin{equation}
\pi_B={\rm i}\left(\begin{array}{cc}
B_{z\bar z}B_{zz}-B_{zz}B_{\bar z z} & {\bf 1}_n+B_{zz}B_{\bar z\bar z}-B_{z \bar z}^2\\
-{\bf 1}_n+B_{\bar z z}^2-B_{\bar z\bar z}B_{zz} & B_{\bar z\bar z}B_{z\bar z }-B_{\bar z z}B_{\bar z \bar z}
\end{array}\right).
\label{trixsz}
\end{equation}
Now the $B$--transform of the $\star$--product (\ref{holstern}) is
\begin{equation}
(f\star_B g)(\bar z, z)=f(\bar z, z)\,\exp\left(-\frac{1}{2}\overleftarrow\partial\pi_B\overrightarrow\partial\right)\,g(\bar z, z).
\label{holbstern}
\end{equation}
It follows that, when  $f,g\in H(\mathbb{C}^n)$, the invariance property (\ref{devs}) does not necessarily hold for arbitrary choices of $B$. However there are some exceptions. For example, when $B_{\bar z\bar z}=0=B_{zz}$ in (\ref{xyzt}), we have a block--antidiagonal, $B$--transformed Poisson tensor
\begin{equation}
\pi_B^{(0)}={\rm i}\left(\begin{array}{cc}
0 & {\bf 1}_n-B_{ z \bar z}^2\\
-{\bf 1}_n+B_{\bar z z}^2 & 0
\end{array}\right)
\label{wenn}
\end{equation}
which, substituted in eqn. (\ref{holbstern}), continues to satisfy the invariance property (\ref{devs}). In general, a necessary and sufficient condition for eqn. (\ref{devs}) to remain valid after a $B$--transformation is the vanishing of the diagonal block entries in (\ref{trixsz}):
\begin{equation}
B_{z\bar z}B_{z z}-B_{zz}B_{\bar z z}=0=B_{\bar z \bar z}B_{z\bar z}-B_{\bar z z}B_{\bar z \bar z}.
\label{allgemein}
\end{equation}
We conclude that, except when condition (\ref{allgemein}) holds, the $B$--deformed product $\star_B$ is now capable of deforming the pointwise product of holomorphic functions. At least in the weak--field approximation made in (\ref{trn}),  and presumably to all orders in $B$, the invariance property (\ref{devs}) will no longer be true. This new feature was absent in the picture in which phase space was regarded merely as a symplectic manifold. It arises as a consequence of the possibility, present in generalised complex manifolds but absent otherwise, of performing $B$--transformations. By their very definition, $B$--transformations are {\it not}\/ diffeomorphisms of the underlying manifold \cite{GUALTIERI}. This fact is reflected in the nontensorial transformation law (\ref{megatrans}). Rather, $B$--transformations correspond physically to the application of external magnetic fields. As in eqns. (\ref{ccw}), (\ref{ccwx}), we may graphically depict this deformation of commutative $C^{\star}$--algebras into noncommutative ones by means of $\star_B$:
\begin{equation}
C^{\infty}(\mathbb{R}^{2n})\stackrel{\star_B}{\longrightarrow}C^{\star_B}(\mathbb{R}^{2n}),
\label{ccwbb}
\end{equation}
\begin{equation}
H(\mathbb{C}^{n})\stackrel{\star_B}{\longrightarrow}H^{\star_B}(\mathbb{C}^{n}).
\label{ccwxbb}
\end{equation}
The new $C^{\star}$--algebra  $H^{\star_B}(\mathbb{C}^{n})$ is noncommutative with respect to the operation (\ref{holbstern})  unless eqn. (\ref{allgemein}) holds.

\section{Discussion and outlook}\label{aussicht}

There are several useful techniques to construct noncommutative $C^{\star}$--algebras \cite{THIRRING}. Two popular ones are the application of background $B$--fields \cite{CDS, SW} and the superposition of branes \cite{WITTEN, BRANES}. Before the advent of generalised complex structures, only Poisson manifolds, but not complex manifolds, could be deformed--quantised. Whenever it was possible to endow a Poisson manifold with a compatible complex structure, the $C^{\star}$--algebra of holomorphic functions remained undeformed (and, in particular, commutative) under the $\star$--operation; this is represented graphically in eqn. (\ref{ccwx}).

In the framework of generalised complex manifolds, we have shown that not only does the $C^{\star}$--algebra $C^{\infty}(\mathbb{R}^{2n})$ of smooth functions  get deformed--quantised, but also the $C^{\star}$--algebra $H(\mathbb{C}^{n})$ of holomorphic functions as well. This conclusion is depicted graphically in eqns. (\ref{ccwbb}) and (\ref{ccwxbb}). A key element in proving our point was the application of $B$--transformations. This latter feature was absent in the usual theory of Poisson manifolds and complex manifolds, where $B$--fields had to be put in externally by hand. On the contrary, $B$--transformations are (nondiffeomorphically realised) symmetries of almost complex manifolds; as such these symmetries are naturally built in.

The previous conclusions have a bearing on coherent--state quantisation \cite{PERELOMOV}. Indeed coherent states are quantum states whose properties are almost classical. Applying $B$--transformations on phase space will induce further {\it quantumness}, the latter being the property opposed to {\it classicality}. These facts are also in agreement with those of refs. \cite{IJMPA, NOI}, where the possibility of quantum--mechanical dualities was examined in the light of generalised complex structures and gerbes.  Specifically, in the terminology of refs. \cite{IJMPA, NOI},  dualities now appear as $B$--transformations. Obviously this relation with coherent states deserves further attention.

Other possible directions for future work are the following. The relation with the quantum--mechanical equivalence principle \cite{MATONE} remains to be elucidated. We have only considered {\it linear}\/ spaces and worked in the {\it weak--field}\/ approximation; nonflat spaces and arbitrarily strong $B$--fields should be analysed. Let us finally add that the {\it industrialisation}\/ of $\star$--products first announced in \cite{LIZZI} can now expand into a new, unexplored market: that of generalised complex geometry.

After completing this manuscript we became aware of ref. \cite{PESTUN}, where issues partially overlapping with those treated here are dealt with.

\vskip1cm
\noindent
{\bf Acknowledgements} It is a great pleasure to thank Albert--Einstein--Institut (Potsdam, Germany) for hospitality during the preparation of this article. This work has been supported by Ministerio de Educaci\'{o}n y Ciencia (Spain) through grant FIS2005--02761, by Generalitat Valenciana, by EU FEDER funds, by EU network MRTN--CT--2004--005104 ({\it Constituents, Fundamental Forces and Symmetries of the Universe}), and by Deutsche Forschungsgemeinschaft.

\end{document}